\documentclass[letterpaper,twocolumn,10.5pt]{article}
\usepackage{usenix2019_v3}

\usepackage{tikz}
\usepackage{amsmath}
\usepackage{titlesec}
\usepackage{xurl}

\usepackage{filecontents}

\newcommand{\odoh}{{ODoH} }

\newcommand{\veryshortarrow}[1][4pt]{\mathrel{%
   \hbox{\rule[\dimexpr\fontdimen22\textfont2-.4pt\relax]{#1}{.6pt}}%
   \mkern-4mu\hbox{\usefont{U}{lasy}{m}{n}\symbol{41}}}}

\usetikzlibrary{positioning, shapes.geometric}

\begin{document}

\date{}

\title{\Large \bf Oblivious DNS over HTTPS (ODoH): A Practical Privacy Enhancement to DNS}

\author{
  Sudheesh Singanamalla $^{\dagger*}$, Suphanat Chunhapanya $^{*}$, Marek Vavruša $^{*}$, Tanya Verma $^{*}$, Peter Wu $^{*}$ \\ Marwan Fayed $^{*}$, Kurtis Heimerl $^{\dagger}$, Nick Sullivan $^{*}$, Christopher Wood $^{*}$ \\
  $^{*}$ Cloudf{l}are Inc., $^{\dagger}$ University of Washington
} 

\maketitle


\begin{abstract}
The Domain Name System (DNS) is the foundation of a human-usable Internet, responding to client queries for hostnames with corresponding IP addresses and records. Traditional DNS is also unencrypted, and leaks user information to network operators.
Recent efforts to secure DNS using DNS over TLS (DoT) and DNS over HTTPS (DoH) have been gaining traction, ostensibly protecting traffic and hiding content from on-lookers. However, one of the criticisms of DoT and DoH is brought to bear by the small number of large-scale deployments (e.g., Comcast, Google, Cloudflare): 
DNS resolvers can associate query contents with client identities in the form of IP addresses. Oblivious DNS over HTTPS (ODoH) safeguards against this problem. 
In this paper we ask \textit{what it would take to make ODoH practical?} We describe ODoH, a practical DNS protocol aimed at resolving this issue by both protecting the client's content and identity. We implement and deploy the protocol, and perform measurements to show that ODoH has comparable performance to protocols like DoH and DoT which are gaining widespread adoption, while improving client privacy, making ODoH a practical privacy enhancing replacement for the usage of DNS. 
\end{abstract}

\section{Introduction}
\label{sec:introduction}


The Domain Name System (DNS) 
is the human bridge to the Internet. In exchange for a human-readable query for a hostname, DNS returns machine-readable addresses and associated records~\cite{borgolte2019dns}. This exchange is handled by resolvers that accept queries from clients and return responses from authoritative name servers. By default DNS messages are transmitted in cleartext using the User Datagram Protocol (UDP) on port 53 (known as Do53). As a result, Do53 is vulnerable to eavesdropping and modification by both well-intended and malicious 3rd parties. These vulnerabilities result in compromising privacy, denial of service (DoS), and injection attacks~\cite{zhu2014t}. In the broader Internet ecosystem where traffic is increasingly safeguarded by HTTPS, Do53 exposes the nature of that activity.

Secure variants of DNS which have recently been introduced can fill that gap. DNS-over-TLS (DoT) \cite{hu2016specification} and DNS-over-HTTPS (DoH) \cite{hoffman2018dns} are widely supported by browsers and increasingly by operating systems. In DoT and DoH, transport of DNS messages between a client's stub resolvers and its upstream recursive resolver are encrypted. However, at time of writing, DoT and DoH are supported only by a small number of large providers (notably Cloudflare, GoogleDNS, and Quad9~\cite{DNSoverHTTPSCloudflare, DNSoverHTTPSGoogle, DNSoverHTTPSQuad9}). Despite the availability of these encrypted protocols and publicly available services, we observed that 92\% of the daily requests (254 Billion/day) received by a popular large recursive DNS resolver use the default cleartext Do53 protocol~\cite{anon} and therefore leak user information.
Although DoT and DoH safeguard DNS with respect to network transit on-lookers, the perceived centralization of DNS resolution brings to bear a secondary concern: While any individual website or online service can associate a request (and user data) with a client and their IP address, DNS resolvers can observe \emph{all} of the client's requested websites and services. 

One potential resolution to this issue is to rely on and trust formal agreements. Mozilla, for example, defined a set of criteria around data retention, aggregation, and sale or transfer, that must be met before a DoH service can be configured in the Firefox browser~\cite{FirefoxDoH, mozillaTRRProgram}. Operational limits on the duration or types of data that might be stored may lend confidence to the safeguarding of users' data. However, in this approach, users' trust must shift from the technical domain to those governed by policy or contracts. While such agreements ostensibly provide privacy, they are difficult \& time-consuming to validate and lack means of enforcement. Moreover the resources necessary to agree and implement such policies are available only to a few organizations that can afford them. 
Lastly, governmental policies and legal decisions may shift, changing already decided agreements. 

Another approach can be architected within DNS itself by ensuring that the resolvers are not given  
client IP addresses. Oblivious DNS (ODNS) was the first protocol that explored technical mechanisms to decouple client IP information from DNS queries. Using a novel recursive resolution algorithm and public key cryptography, ODNS ensures that no single entity other than the client is privy to both the IP address and DNS query~\cite{schmitt2019oblivious_full}. The ODNS protocol is designed to be compatible with existing DNS architectures and introduces ODNS resolvers that operate between the recursive resolvers and authoritative name servers. The recursive resolvers receive an encrypted query and forward it to the intended ODNS resolver which decrypts and retrieves a response from the authoritative name server. When compared to Do53, the ODNS architecture significantly impacts performance since every query is destined to an authoritative name server instead of more performant recursive resolvers. It also doesn't completely protect clients: Recursive resolvers can still supply clients' IP addresses using EDNS0 to both a colluding ODNS server and the authoritative name servers. 
Recent efforts in IETF have emerged to adopt Oblivious DNS ideas into DoH~\cite{odoh-draft}~\footnote{We note for clarity of contribution that there is an overlap between the participants in the IETF drafting process and this work.} The addition of HTTPS layer on top of ODNS solves the problem of establishing server identity through TLS and guarantees in-order delivery of packets due to TCP preventing the need for any state machines in ODNS. 


In this paper we build upon these prior efforts by architecting, implementing, and deploying multiple interoperable instantiations of Oblivious DNS over HTTPS (ODoH). To the best of our knowledge these are the first and only implementations of the standard. We use them to answer a number of questions: (1) What does it take to deploy a large scale ODoH infrastructure? (2) Where are the performance gaps or penalties? (3) How do our experiences implementing and evaluating this protocol inform our suggestions for practical global adoption? We answer these questions by comparing the protocol to other available encrypted DNS protocols, like DoT and DoH that provide similar privacy guarantees, and default cleartext usage of DNS. We find that ODoH has DNS response time performance which is better than other available protocols offering the same privacy guarantees. We also find that the usage of ODoH incurs a  minimal page load time performance impact compared to other DNS protocols.


\section{Background \& Related Work}
\label{sec:background_related_work}

The Domain Name System (DNS) was initially proposed as an IETF standard using UDP to help clients lookup the IP address of the destination server that the client wants to communicate with. The clients send a UDP DNS packet to the DNS recursive resolver, a server which receives a client query and resolves the query to a corresponding IP address, using \textit{cleartext} over a computer network. Therefore, the standard communication protocol for DNS using cleartext UDP packets (Do53) does not have any confidentiality or integrity guarantees, making such communication vulnerable to eavesdropping and tampering attacks~\cite{bortzmeyer2015dns, bortzmeyer2016dns, morecowbell}. Additionally, attempts made by regulators to require ISPs to actively block access to content on the web (objectionable or through legal notices) have been struck down due to the security risks they pose and violations to free speech. Regulations also introduce the need for the DNS resolvers operators to provide parental controls which use DNS filtering to restrict access~\cite{ISPAMozillaVillain19, spence2005pennsylvania, cleanbrowsing}.


To overcome the challenges due to cleartext communication, various techniques like DNS-Over-TLS (DoT), DNS-Over-HTTPS (DoH)~\cite{hoffman2018dns}, and DNSCrypt have been proposed and implemented enabling encrypted DNS communication between the client and a recursive DNS resolver.  Desktop web browsers~\cite{FirefoxDoH, ChromeDoH, ChromiumDoHBlog}, mobile clients~\cite{AppleWWDCDoH, AppleDNSProxy, AndroidDNSDoT} and operating systems~\cite{MicrosoftDoHInsider} have recently started gaining supported to use DoH. DoH and DoT protocols rely on a TLS session between a client and the resolver providing clients with an authenticated and secure connection to the DNS resolver. The usage of DoH has been increasing across the web due to public services offered by Cloudflare and NextDNS and their integration within Firefox as the trusted recursive resolver with Cloudflare being the default~\cite{FirefoxDoH, lu2019end, bottger2019empirical}. While organizations like Google and Cloudflare operating DoH resolvers guarantee privacy through policies which prevent user tracking such as periodically (every 24 hours) purging all user traffic data~\cite{CloudflareDNSAudit}, some organizations do aggregate user traffic patterns which could be beneficial to their own businesses~\cite{cloudflareprivacy, DNSoverHTTPSGoogle}.

\subsection{Prior Encrypted DNS Protocols}
\label{ssec:prior_dns_encryption_protocols}

Popular encrypted DNS protocols which have gained significant adoption (DoT, DoH)  involve encrypting the network channel and sending a plain-text message through the channel. DNSCrypt~\cite{DNSCrypt} and its predecessor DNSCurve~\cite{bernstein2009dnscurve} are protocols which encrypt the DNS message and are sent as TCP or UDP packets. A client connecting to the DNSCrypt resolver receives a public set of signed certificates which is verified by the client using a known \textit{provider} public key. Each certificate contains a short-term \textit{resolver} public key. The client uses the resolver public key and the client secret key and uses an agreed upon key exchange algorithm defined in the certificate to generate a shared key with which subsequent queries are encrypted using an authenticated encryption algorithm. The client sends the encrypted messages using UDP or TCP to the intended DNSCrypt resolver over a non-encrypted channel. Both the channel and  message based encrypted DNS protocols pose privacy risks where the operator of the resolver could log and store information about the client allowing them to monetize the user traffic and create a profile of the client.

Anonymous DNSCrypt attempts to address the issue by introducing public non-logging proxies through which the DNSCrypt requests could be routed to the intended DNSCrypt resolver~\cite{AnonymizedDNSCrypt, AnonDNSCryptImplementation}. While it is possible to achieve the intended privacy goals through the usage of DNS-over-HTTPS over the Tor network (DoHoT)~\cite{DNSoverTorCloudflare, DoHoTPractical}, the layered encryption and three node default circuits specification add significant performance bottlenecks for clients to obtain the answer from the resolver. Many countries have been actively implementing techniques to censor Tor nodes and their traffic and block them from the Internet~\cite{TorBanUAE, TorBanIran}. In addition to prior successful DNS fingerprinting attacks over Tor~\cite{greschbach2016effect}, the volunteers running the exit nodes through which the client traffic flows through face significant legal liabilities~\cite{TorEFFMisunderstanding, TorSeattleCapture}. With ODoH we aim to provide the privacy guarantees by bringing together popular non-colluding DNS service providers and volunteers with a shared purpose of improving privacy. Based on our evaluation, we detail a practical roadmap for the same in Section~\ref{ssec:roadmap}.

\subsection{Privacy \& Regulatory Considerations}
\label{ssec:privacy_in_dns}

DNS in the existing encrypted variants discussed in Section~\ref{ssec:prior_dns_encryption_protocols}, and the cleartext UDP variant reveal significant information about the client IP addresses, allowing operators of these resolvers to geolocate their clients. Previous research has shown that the Tor approach can result in leakage of client information and the ability to identify users when visiting a hostname through the anonymized service based on the DNS lookups~\cite{greschbach2016effect}. With DNS over HTTPS over Tor (DoHoT)~\cite{DoHoTPractical}, the Tor network paved a new way in which users readily achieved privacy of their DNS queries by moving clear-text DNS queries to DoH queries sent through the Tor network. However, this approach resulted in a significant performance impact for users in both the DNS response times and web page load times. We also identify that Internet Service Providers (ISPs) operating recursive DNS resolvers configure themselves as the default DNS resolver to be used by the client when granting an IP addresses using the DHCP protocol enabling the ISPs to access every query made by the client on the network. The collection of such queries can paint a revealing picture of the subscribers' habits and interests~\cite{ISPView}. Additional concerns have been raised in prior work due to the non-consensual approaches of ISPs assigning DNS configurations for their clients requiring users to manually change their DNS providers to a different one that they trust~\cite{borgolte2019dns}. The Federal Communications Commission (FCC) raised significant concern over an ISP's ability to view their subscribers' DNS queries in cleartext and heavily regulated any attempts at monetizing the DNS data~\cite{federal2016protecting}. 

A large number of open resolvers eg. Cloudflare, Quad9, NextDNS and Google DNS provide recursive DNS resolver services which can be configured by the clients as an alternative to the default DNS resolver configured by their ISP provider. However in the process of changing the default provider, the user implicitly places trust in the chosen recursive resolver. This does not alleviate the privacy concerns since the resolver being used can still collect and store the client request information thereby being able to create a profile about their users. Changing the default ISP issued DNS provider to a trusted DNS resolver operator only shifts the privacy problem from ISPs being able to profile the user to a trusted operator being able to do the same. While trusted services promise to purge all user traffic information periodically (eg. every 24 hours), there is a need to enable privacy technologically through a newer protocol than legally through a strict privacy policy~\cite{cloudflareprivacy}. Some ISPs providing DNS resolver services indicate that only error traffic is monetized~\cite{weaver2011redirecting}. In a critique, Borgolte \textit{et al.} examine the performance of unencrypted DNS with DoH and critique that the increase in encrypted DNS traffic ultimately enables a single organization or a service provider to observe and monetize DNS traffic and improve their efficiency to deliver targeted advertising~\cite{borgolte2019dns}. Additional criticism regarding centralization of the DNS for the web along with the increased switching costs for consumers to exercise their choice of provider due to browser/OS tie-ins raise additional regulatory and policy implications with regard to net neutrality~\cite{borgolte2019dns}. Recent efforts by Mozilla to provide users with more options resulted in negotiations with ISPs like Comcast 
to provide DoH services conforming  to a strict privacy policy and rules of the TRR program~\cite{mozillacomcast20, mozillaTRRProgram}.

\subsection{Attempts at DNS Privacy}
\label{ssec:oblivious_variants_dns}

Prior work by Zhao \textit{et al.} proposed an approach of \textit{range queries} where random noise, and the usage of Privacy Information Retrieval (PIR) techniques, was introduced during the execution of the DNS queries which when used in conjunction with DNSSEC aims to provide confidentiality, integrity and privacy protections~\cite{zhao2007analysis, zhao2007two}. However, an evaluation by Castillo-Perez \textit{et al.} of these two protocols indicated that despite improvements since the first proposal in reducing bandwidth usage with PIR techniques, there were serious feasibility issues since the techniques involved changes to both the DNS servers and the DNS clients~\cite{castillo2009evaluation}. Additionally, the evaluation also indicated serious security flaws where the attackers who can control the channel intercepting the ranges of the packets can infer the original query defined by the user and efficiently forge a response~\cite{castillo2009evaluation}.

T-DNS leverages connection oriented TCP with the addition of pipelining DNS queries and performing Out-of-order processing (OOOP) at the recursive resolvers to achieve the performance similar to UDP based DNS protocols which are in widespread use and position the usage of encryption as a privacy enhancement when comparing to clear-text UDP based DNS protocol~\cite{zhu2014t}. Prior approaches leveraged cryptographic mixes / mix cascades in to anonymize the user traffic which is a model widely adopted in Tor~\cite{federrath2011privacy, berthold2001web}. Additional approaches involved broadcasting traffic of a small fraction of hostnames which might cover the vast majority of all user traffic making it more difficult to thwarting attempts at profiling~\cite{federrath2011privacy}. However, such approaches face practical limitations due to large bandwidth usage and long-tailed distribution of queries which ensued a hybrid approach of using both mix cascades and broadcasts~\cite{federrath2011privacy}, but suffers from degraded performance due to increased page load and DNS response times.

\subsection{Prior Measurements}
\label{ssec:prior_measurement}

The large scale measurement of DNS over Encryption protocols indicated that the adoption of usage of encryption in DNS queries has been increasing due to the implementation of support for such services in operating systems and by public open DNS resolvers~\cite{lu2019end}. B{\"{o}}ttger \textit{et al.} in their work on measuring the cost of DNS-over-HTTPS (DoH) conclude that a switch to DoH does not significantly impact page load times and does so without sacrificing user security~\cite{bottger2019empirical}. Sundaresan \textit{et al.} perform large scale measurements from home networks to identify performance bottlenecks and indicate that as broadband access speeds continue to increase, latency becomes a performance bottleneck with metrics like DNS response time and time to first byte becoming more important~\cite{sundaresan2013web}. Sundaresan \textit{et al.} also mention that the page load times can be improved by prefetching~\cite{sundaresan2013web}. Recent efforts by Hounsel \textit{et al.} conclude that the performance of encrypted protocols (DoT, DoH) vary by the choice of the public DoH resolver~\cite{hounsel2020measuring}. In our measurements with ODoH, we choose three popular public DoH resolvers and randomize our queries to them from the oblivious target in an effort to reduce measurement biases because of usage of one specific resolver.



\section{Oblivious DNS over HTTPS (ODoH)}
\label{sec:odoh}

At a high level, the design of Oblivious DNS over HTTPS (ODoH) is similar to that of DNS over HTTPS (DoH) with the addition of an intermediate proxy node which performs a query on behalf of a client. However, there are many differences which enable the obliviousness of the protocol compared to a proxied variant of DoH (pDOH) where the DoH request is delegated to the proxy instance by the client. The broad mandate of ODoH is to prevent recursive resolvers, and ISPs running such resolvers, from being able to successfully link clients to their requests. As such, the aim of ODoH is to (1) encrypt DNS queries being sent to the DNS resolver, and (2) remove the need for the resolver to observe client IP addresses to respond. ODoH improves upon the limitations of ODNS and can also support 0x20 bit encoding which is commonly used for DNS forgery resistance~\cite{dagon2008increased, schmitt2019oblivious_full}.

\subsection{Security, Trust \& Threat Model}
\label{ssec:security_assumption_threat_model}

\textbf{Assumptions: }ODoH assumes a Dolev-Yao style attacker who can monitor and observe all the requests on a specific network channel i.e. between the stub resolver and the proxy or proxy and the target~\cite{dolev1983security}. Each attacker has complete control over a specific network channel and can control the flow and content of the request but do not collude with the target resolvers or other attackers on different network channels. Our goal is to ensure that the ODoH protocol achieves confidentiality guarantees in the presence of such an attacker.

\textbf{Guarantees:} ODoH guarantees that the queries made are known only by the client stub resolver initiating the query and the intended target resolver holding the corresponding private key to decrypt the message. At any point in the execution, oblivious clients know the query they made, the intended target, and the proxy chosen and can verify if the returned response has been encrypted correctly. The oblivious proxy knows the client stub resolver IP address and the target chosen by the client, but cannot recover any information about the query being made. The oblivious target knows the IP address of the proxy forwarding the encrypted query and can decrypt the message to obtain the plaintext DNS question. The target resolvers cannot link the queries to the actual client making the queries unless a client stub resolver reuses the key using which the target needs to encrypt the response.


\begin{figure}
	\centering
	\includegraphics[width=0.9\columnwidth]{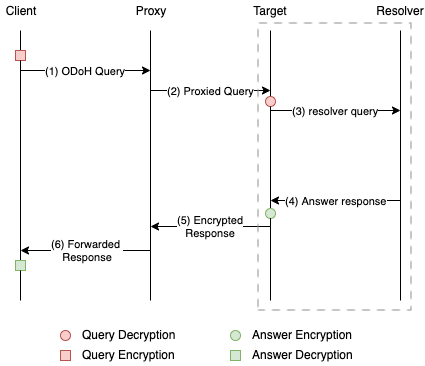}
	\caption{ODOH Protocol Execution Sequence Diagram}
	\small{The sequence of network requests between various components in ODoH. The dashed box indicates possible component co-location.}
	\label{fig:odoh_protocol_execution_sequence}
\end{figure}



\subsection{ODoH Design}
\label{ssec:odoh_components}

In this section we introduce the \odoh design, including (1) an Oblivious Target that is ideally co-located with the recursive resolver, (2) an Oblivious Proxy 
through which (3) clients communicate using a client stub resolver. We also detail the implementations of these components which are later used for measurements.

The \textbf{Oblivious Target} is a hosted instance that receives encrypted DNS queries, decrypts and resolves them, then encrypts and returns the response to the client. The oblivious target publishes a public key which the clients securely retrieve and use to encrypt their DNS queries. The target can be coupled with the recursive resolver to improve performance.


The \textbf{Oblivious Proxy} is a hosted instance that receives the encrypted DNS query containing the DNS request, as well as the intended oblivious target from the client. The proxy removes information about the client (such as their IP address) and forwards the request to the chosen oblivious target. The oblivious proxy then receives the encrypted DNS response from the oblivious target, returning it to the client.


\textbf{Clients} using the \odoh protocol run a modified \odoh stub resolver, replacing their default Do53 DNS usage. The client stub resolver can be configured to use a set of chosen trusted oblivious proxies and oblivious targets or can be configured to discover and choose randomly between a known set of proxies and targets. The client stub resolver encrypts the DNS query and requests the oblivious proxy to forward the encrypted request to the target resolver.

\textbf{Network Architecture: } 
\label{ssec:odoh_architecture}
We wish to specifically emphasize that ODoH introduces new layers into the DNS hierarchy, such as oblivious proxies and oblivious targets, increasing the total number of network transits necessary to obtain a DNS response. It adds two network requests, the proxied request from the proxy to the resolver and response from the recursive resolver to the proxy. It is possible that the oblivious target would similarly add two transits, though it can instead be implemented directly into a DoH recursive resolver or as a standalone server which communicates with the recursive resolver. We examine both cases in our evaluation. 

\subsection{Protocol Execution}
\label{ssec:odoh_protocol_execution}

We present the execution of the protocol as a sequence diagram in Figure~\ref{fig:odoh_protocol_execution_sequence}. The sequence diagram indicates the target and the resolver as separate components but could be co-located at the resolver in practice. We discuss the impact of performance due to such co-location in section~\ref{ssec:colocation_impact}.


\textbf{Discovery in ODoH \& Verification:}
\label{sssec:odoh_discovery}
Traditionally, DNS configurations are issued directly by ISPs during DHCP or are manually configured by the user. For ODoH, similar to DNSCrypt or DoH, it is possible for an interested party to host a centralized list of proxy and target server hostnames which support the ODoH protocol. Such a list could be bootstrapped in browsers, like Firefox or Chrome, and operating systems which aim to support the protocol. Each client stub resolver requests a DNSSEC signed resource record set (HTTPSVC or SVC or similar special fields) where the ODoH public key for the target is published. The stub resolver retrieves this information and stores it after validating the signature using the public key from the DNSKEY records in addition to validating the chain of trust to ensure that the DNS record is retrieved from the authoritative server and was not altered en-route. Once validated, the stub resolver adds the oblivious target to the list of verified targets and stores the public key necessary for communication. These records can be periodically validated and can result in corresponding updates to the configuration of the client stub resolver. When a browser or a client joins a network, a special lookup is made for \texttt{odoh.test} which returns the necessary keys for validation and indicates the support for ODoH on the resolver. This is similar to how \texttt{doh.test} is currently used to check if a resolver supports DoH~\cite{mozillacomcast20}. This is particularly chosen as the discovery and key distribution mechanism to reduce barriers to practical real world deployments.

\textbf{ODoH Query Request:}
\label{sssec:odoh_query_request}
ODoH uses a Hybrid Public Key Encryption (HPKE) scheme~\cite{barnes2019hybrid}. The client stub resolver prepares the DNS query and generates a symmetric key and encapsulates the key and a DNS query into a message which is then encrypted using the public key of the target/resolver obtained from the discovery and verification phase of the protocol. The stub resolver sends out the request in the DNS wire format and introduces a new content-type \texttt{application/oblivious-dns-message} which is used to send the message to the chosen oblivious proxy along with forwarding instructions to the appropriate target.

\textbf{ODoH Answer Response:}
\label{sssec:odoh_answer_response}
The proxied query is received by the intended target resolver which decrypts the encrypted query and decapsulates the message to obtain the symmetric key sent by the client along with the DNS query. The query is resolved appropriately, either due to a cache hit on the co-located resolver or by performing a request between the target and the resolver to obtain an answer. The answer response is encrypted using the client provided symmetric key. The encrypted answer is sent to the proxy as a response who forwards to the client that issued the query. The client then decrypts the obtained response using the symmetric key.

\subsection{Implementation}
\label{ssec:implementation}

\textbf{Target and Proxy:} We implemented two interoperable variants of the oblivious target and proxy using both Golang and Rust in an effort to prepare for the IETF standardization process~\cite{housley2011reducing} and hosted them  on two cloud platforms: Google Cloud using Google App Engine and Cloudflare workers~\cite{CloudflareWorkers}.

\textbf{Client:} We implemented a Golang based client with a command line API similar to \texttt{dig} which uses the ODoH protocol and performs queries to a chosen oblivious proxy and oblivious target. We implemented a benchmarking submodule which we use for evaluations which is capable of instantiating \textit{C} client processes, each performing \textit{N} queries at a rate of \textit{R} DNS requests per minute. Our implementation supports a wide number of cryptographic cipher suites which can be used for encryption and decryption of the DNS queries and answers. Our implementation also provides the ability to reuse HTTPS connections to avoid new TCP and TLS handshakes, which is enabled for measurements in this paper. We present a comparison indicating the impact of connection re-use later in section~\ref{ssec:connection_reuse_impact}. In this work, we implemented the oblivious target and clients with the support of the HPKE scheme using X25519 Key Exchange Mechanism (KEM), SHA256 based Key Derivation Function (KDF) and AES-128-GCM based Authenticated Encryption (AEAD) algorithms. Additionally, we support X448, P256, P521 as available KEMs; SHA384, SHA512 as available KDFs; AES-256-GCM, ChaCha20Poly1305 as the authenticated symmetric encryption schemes. For the evaluations of ODoH, we use ODoH clients which use the discovery lists, retrieve and validate the public keys of the targets, and measure the latency of different proxy-target pairs and decide to use the pair with the lowest latency. We present our evaluation methodology in section~\ref{sec:measurement_methodology} and present our results in section~\ref{sec:results}.

We also integrated ODoH protocol support into popular open source client stub resolvers which support DoH or encrypted DNS protocols like DNSCrypt. We deployed the client stub resolver on a local machine to perform DNS lookup and page load time measurements discussed in section~\ref{ssec:page_load_impact}. We have open sourced all of our interoperable implementations of the proxy, target \& clients supporting the ODoH protocol~\footnote{Links to the code are anonymized during the review of this paper}.






\section{Measurement \& Methodology}
\label{sec:measurement_methodology}

We perform measurements by deploying the oblivious proxies and targets through  geo-replicated serverless platforms. We also provision 9 VMs, one in each of the seven public Google Cloud data centers in the United States, one in Canada and one in South America. Each of these geolocated data centers has the serverless proxy and target instances in the nearest datacenter to them on the Cloudflare network. All the VMs chosen for the experiment have 1 Intel Xeon \@2.0 GHz CPU core with 3.75 GB of memory and use x86\_64 architecture~\cite{gcecloud} with an average bandwidth of 480 Mbit/s. Additionally, we host two scalable but geographically fixed Google app engine deployments for the proxy and the target instances in a Google Cloud datacenter on the west coast of the USA.

During the experiment, we use 90 client stub resolvers, where 10 clients are run in parallel on each of the VMs. Each client stub resolver makes 200 ODoH queries through the experiment at a rate of 15 requests/minute to create a stress similar to ($\approx$21000+ DNS requests/day) issued by a client eg. browser/laptop/phone as measured by capturing wireshark packet traces on the clients ~\cite{DNSqueryAveragePiHole}. 

During the experiments, we deterministically choose the recursive resolver from the oblivious Target instance based on the symmetric key to distribute the client queries across three popular open DoH resolvers (Cloudflare DNS, Google DNS and Quad9). This is intentionally done to reduce any biases in measurement due to the cloud operators' network.

We additionally compare the ODoH protocol proposed in this work to existing DNS protocols and their network variants. We establish the baseline benchmark using DoH where each client instance directly performs DoH queries to one of the three chosen resolvers randomly. We compare this to a proxied DoH variant (pDoH) where the client stub resolver requests a hosted oblivious proxy to perform the DoH requests. This variant of the protocol architecturally introduces one additional hop between the client and the recursive resolver.

To understand the impact of encryption and decryption compute costs, we compare ODoH where the client sends a request to a proxy which forwards it to a target which queries the resolver, we perform the same queries as DoH without additional encryption/decryption costs. We refer to this no-encryption variant of ODoH as Cleartext-ODoH. In our work, we also compare ODoH to encrypted DNS variants such as DNSCrypt, and DNS over HTTPS over Tor.

We use dnscrypt-proxy~\cite{DNSCrypt, DNSCryptProxyImpl}, an official open source implementation of the DNSCrypt proxy and configure it to use only the DNSCrypt protocol and avoid fallbacks to Do53, DoH. We explicitly disable the usage of TCP in DNSCrypt since the implementation according to the original specification uses DNSCrypt with UDP. dnscrypt-proxy uses the default configuration to force TCP connections instead of UDP and does not reuse connections for privacy reasons. To enable DNSCrypt for its highest performance we disable TCP, create a new ephemeral key for each query to draw parallels with ODoH and allow the stub resolver to choose the nearest resolver with the lowest latency which supports DNSCrypt from the public list of DNSCrypt resolvers available.

We also compare the results of ODoH with DoHoT. For these evaluations, We grant similar benefits to Tor and allow Tor to create its own optimal circuit for guaranteeing privacy and perform DoH queries over Tor using the SOCKS5 proxy on each client instead of configuring them explicitly to use a static route. In the next section we share the results of our experiments and comparisons of various DNS protocols. The various protocols considered for the measurement are also summarized in Table~\ref{table:network_path_variance} for reference, alongside a high level sequence of their path components.

\begin{table*}[h]
\centering\small
\begin{tabular}{|l|l|l|l|l|}
\hline
\textbf{Protocol} & \textbf{Request Path} & \textbf{Security} & \textbf{Privacy} & \textbf{Description} \\ \hline
Plain DNS (Do53)~\cite{mockapetris1987domain}     & C $\veryshortarrow$ R   & No  & No & UDP-based default clear-text DNS         \\ \hline
DNS over HTTPS (DoH)~\cite{hoffman2018dns}               & C $\veryshortarrow$ R   & Yes & No$^{*}$ & HTTPS based DNS over HTTPS \\ \hline
Proxied DoH (pDoH) & C $\veryshortarrow$ P $\veryshortarrow$ R & Yes & No & Proxy performs DoH query\\ \hline
\textbf{Oblivious DoH (\odoh)}~\cite{odoh-draft} & C $\veryshortarrow$ P $\veryshortarrow$ T $\veryshortarrow$ R & Yes & Yes & Oblivious DoH protocol \\ \hline
\textbf{Cleartext ODoH} & C $\veryshortarrow$ P $\veryshortarrow$ T $\veryshortarrow$ R & Yes & No & DoH query is proxied to a Target which performs DoH query \\ \hline
\textbf{Co-located ODoH}   & C $\veryshortarrow$ P $\veryshortarrow$ (T+R)  & Yes & Yes  & ODoH with Colo'd Target and Resolver \\ \hline
DNSCrypt~\cite{DNSCrypt}          & C $\veryshortarrow$ R    & Yes & No$^{*}$  & UDP-based encrypted DNS \\ \hline
Anonymous DNSCrypt~\cite{AnonymizedDNSCrypt}  & C $\veryshortarrow$ P $\veryshortarrow$ R  & Yes & Yes  & UDP based, Proxy routes client query         \\ \hline
DoH over Tor (DoHoT)~\cite{DoHoTPractical, DNSoverTorCloudflare} & C $\veryshortarrow$ Tor $\veryshortarrow$ R & Yes & Yes & DoH over Tor Network \\ \hline
\end{tabular}
\caption{Evaluated DNS alternatives. C - Client, P - Proxy, R - Resolver, T - Target}
\label{table:network_path_variance}
\small{*-The guarantee of privacy depends on the resolver being used. Privacy is guaranteed using legal mechanisms such as privacy policies.}
\end{table*}


\section{Results}
\label{sec:results}

\begin{table}[h]
\begin{tabular}{|l|p{2.5cm}|l|}
\hline
\textbf{Microbenchmark} & Type & P99 \textbf{Overhead} \\ 
\hline
ODoH Encryption & X25519/SHA256 & 360$\mu$s \\
ODoH Decryption & X25519/SHA256 & 246$\mu$s \\
ODoH Query  & AES-128-GCM & 107B \\
ODoH Answer & AES-128-GCM & 16B \\
\hline
\end{tabular}
\end{table}

\subsection{Microbenchmarks}
\label{ssec:microbenchmarks}

\textbf{Cryptographic Compute Overheads:}
\label{sssec:compute_overheads}
To identify compute overheads incurred by the additional Hybrid Public Key encryption and decryption phases of the protocol,
we perform a microbenchmark on the client VM instances at sub-microsecond granularity. 
We perform this benchmark by randomly choosing 10000 domains from the Tranco million dataset~\cite{pochat2018tranco} and encrypting each DNS type A query across a different public key. 
The encryption and decryption operations are executed sequentially.
At the $99^{th}$ percentile the encryption of a DNS query using the most performant cipher suite (X25519 KEM, SHA256 KDF and AES-128-GCM AEAD) takes 360$\mu$s, while the decryption of such an encrypted query takes 246$\mu$s. The protocol also supports other cipher suites involving X448, P256, and P512 KEM; SHA384 and SHA512 KDF; and AES-256-GCM and ChaCha20Poly1305 AEADs. In some cases P256 curve might be preferred over the X25519 KEM for NIST and FIPS standards compliance~\cite{pub1994security}. For the symmetric key operations AES performs better than ChaCha20Poly1305 due to hardware acceleration in some platforms~\cite{AESGCMPerformance}. Curves with a higher bit size like X448 and P521 are desirable for clients that prefer a higher security level at the cost of computation overheads. Our results indicate an extremely minimal compute overhead for oblivious target instances to decrypt the received messages. The client stub resolvers using the ODoH protocol incur marginally higher compute overheads for encryption of the DNS queries. In our evaluations the slowest encryption times at the 99th percentile is 430$\mu$s using the X25519 KEM with different possible permutations of the supported KDFs and AEADs;
decryption is 713$\mu$s when using the SHA 384 and AES 256 GCM. However, all the computation overheads in the lifecycle of an ODoH query, i.e. from query encryption at the client to
decryption of the response,
takes less than 1ms.
The use of parallelism on multi-core machines could further improve performance.


\textbf{Network Overheads:}
\label{sssec:network_overheads}
At the client stub resolver, usage of HPKE encryption 
increases the 
size of the query on the wire by 4x, and the 
and the size of the response by 1.2x, when compared to the baseline clear-text DNS messages. This microbenchmark randomly samples 10000 unique domains and creates a serialized DNS message 
with the query. 
The same set of domains are serialized 
for the corresponding ODoH message. On average, each clear-text DNS query has a wire size of 33.8 bytes.
The average encrypted ODoH query is 314\% larger at 140.8 bytes.
We perform the DNS queries and collect the responses obtained while performing the microbenchmark and compare the average response sizes. The \odoh encrypted responses 
increase the baseline 71.5 bytes size by 22.37\% to 87.5 bytes. 
Note that \odoh messages encapsulate the client public key used to encrypt the query, appended to an integrity hash. 

\subsection{Macrobenchmarks}
\label{ssec:macrobenchmarks}

\begin{figure}[h]
	\includegraphics[width=0.9\columnwidth]{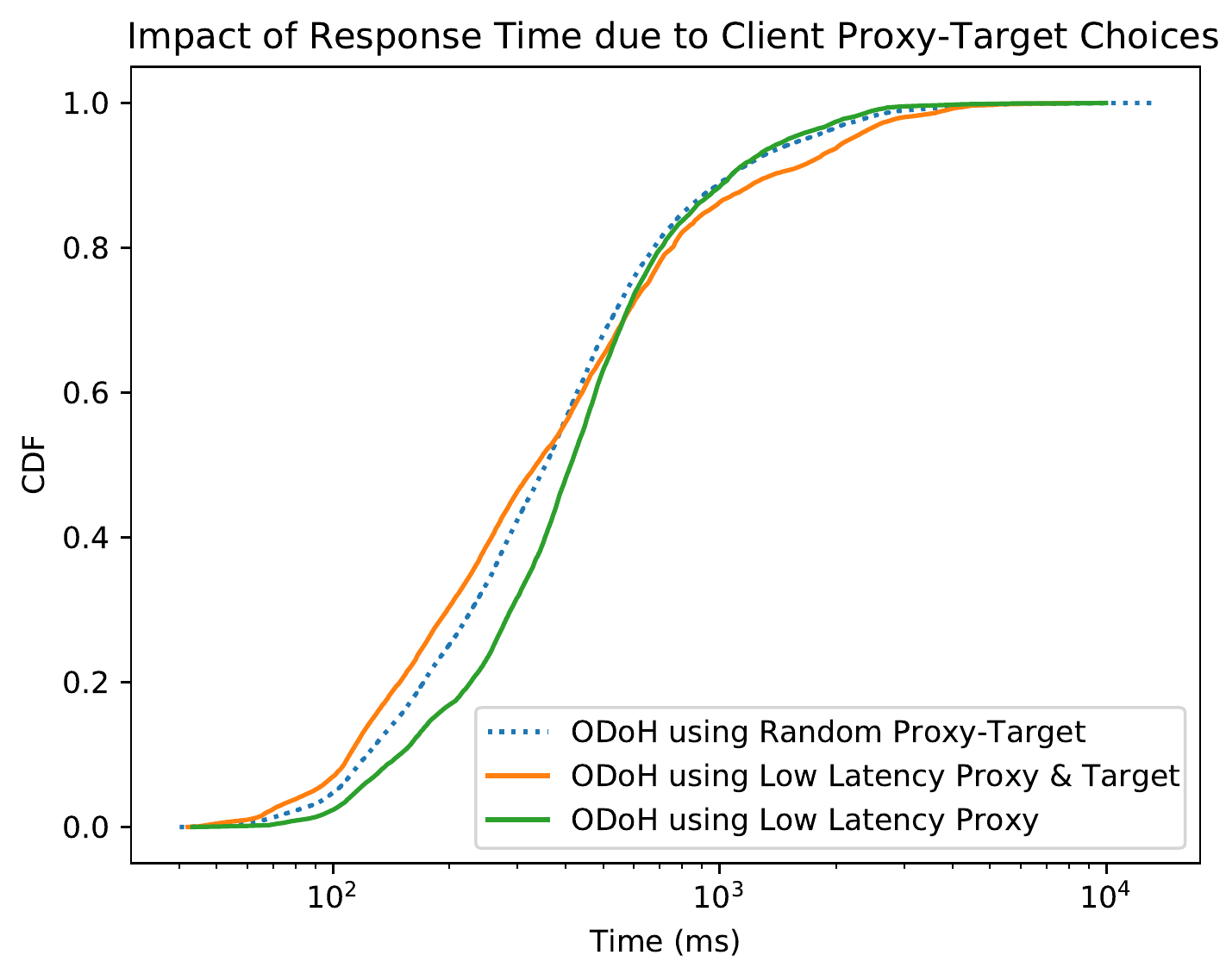}
	\caption{Impact of ODoH due to Proxy-Target selection}
	\label{fig:odoh_variants_compare}
\end{figure}

We evaluate \odoh by arranging its components into three 
configurations: (1) the client stub resolver chooses a random proxy-target pair; (2) the client stub resolver chooses the lowest-latency proxy with a random target; and (3) the client stub resolver chooses the proxy-target pair with the lowest total latency.  
The measured response times are shown in Figure~\ref{fig:odoh_variants_compare}, and indicate that
the choice of proxy and target does indeed 
impact \odoh performance. 
We note that the lowest total latency proxy-target pair 
performs better than choosing only the lowest-latency proxy and random target.
In our evaluations
the average query to response time 
over the lowest latency proxy-target pair is 334.85 ms compared to 411.44 ms when using the lowest latency proxy. This marks an average improvement of 22.8\%
and points to optimal performance when
the proxy is on-path to the intended target, since an on-path proxy should yield 
minimal latency costs (all other factors being equivalent, such as connection reuse potential).

\begin{figure}[t!]
	\includegraphics[width=0.9\columnwidth]{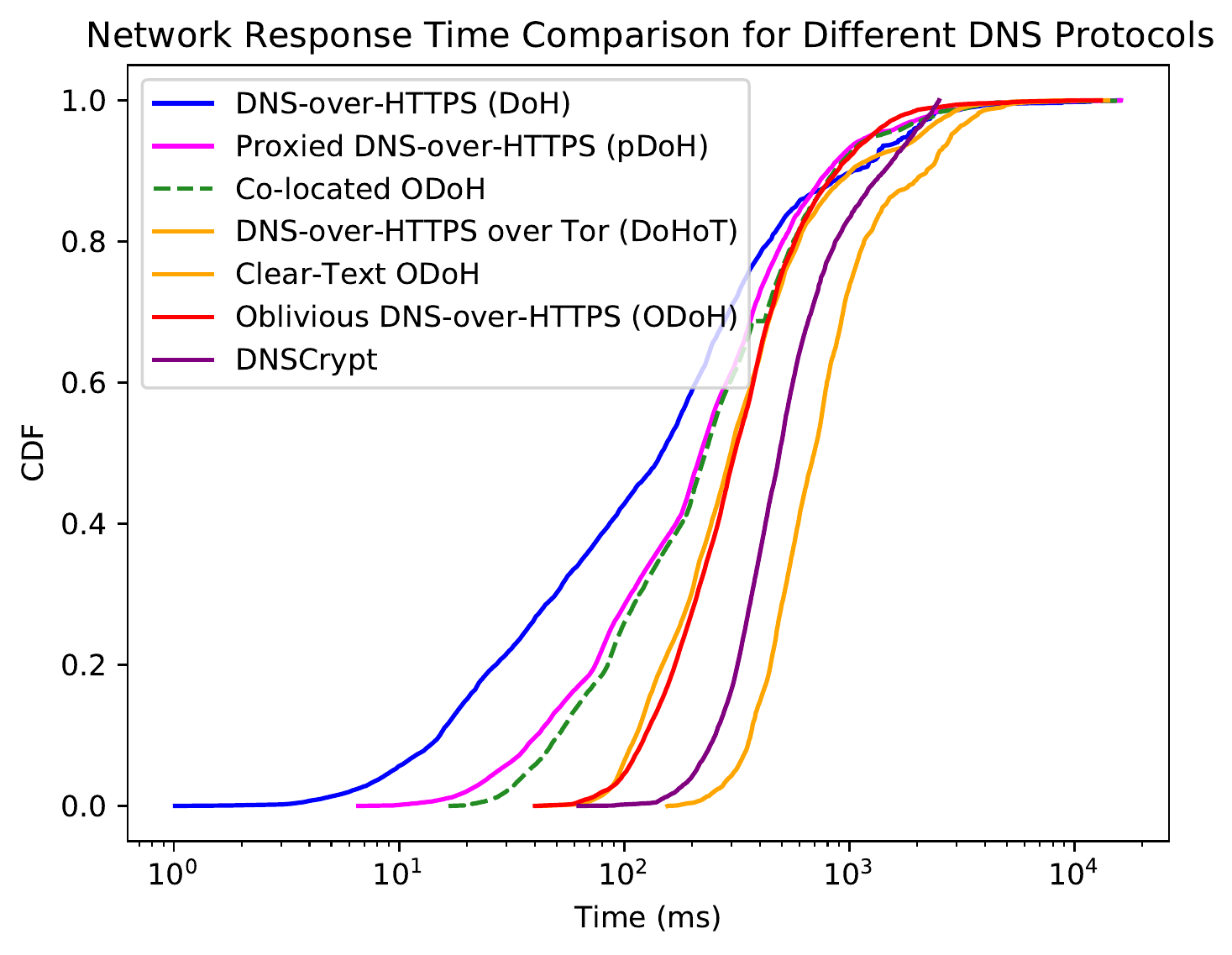}
 	\caption{Comparison of ODoH to other DNS Protocols}
	\label{fig:all_protocols_compare}
\end{figure}

Figure~\ref{fig:all_protocols_compare} shows the CDF of various ODoH instantiations relative to other DNS architectures, as described in Section~\ref{sec:measurement_methodology}. ODoH is the best performing privacy enhancing DNS protocol, ahead of DNSCrypt and DoH over Tor. Also, \odoh performance is only marginally worse than \odoh in clear-text, where a standard DoH query traverses the same logical 3-hop path as an \odoh query, indicating minimal cryptographic compute overheads as presented in the in Section~\ref{ssec:microbenchmarks}.

We use the standard DoH response time as the baseline against which we compare the the various protocols. On an average, we measure that the the DoH response time is 146.12 ms. In the proxied DoH variant where the client delegates the proxy to perform a DoH query directly to the resolver, the additional network request between the proxy and the resolver increases the average response time by 49.5\% to 218.52 ms. Co-located ODoH, architecturally identical to pDoH with the proxy and target in the same place, sees a slight increase in response time because of the decryption and cache operations. We detail our analysis on impacts of co-location in Section~\ref{ssec:colocation_impact}. If the same DoH query is sent through both the proxy and the target, located on different machines, before being resolved by the resolver (clear-text ODoH), we see an increase in average DNS response times by 103.2\% to 297.015 ms which increases to 305.11 ms when using ODoH with the compute overheads of decryption and encryption at the target and decryption of the response at the client. Other privacy enhancing DNS protocols like DNSCrypt perform much worse even when using a resolver with the lowest latency and increase the average response time by 233.7\% compared to DoH to 487.68 ms while DoHoT incurs a 376.5\% average performance overhead when compared to DoH with response times at 696.40 ms. 
Results for the DNSCrypt protocol indicate that the response times through ODoH are 59.8\% faster despite lack of co-location suggesting \textit{atleast} an improvement of 60\% in the query response time of \odoh relative to remaining available privacy preserving options.

\subsection{Impact of Connection Reuse}
\label{ssec:connection_reuse_impact}

\begin{figure}[t!]
	\includegraphics[width=0.9\columnwidth]{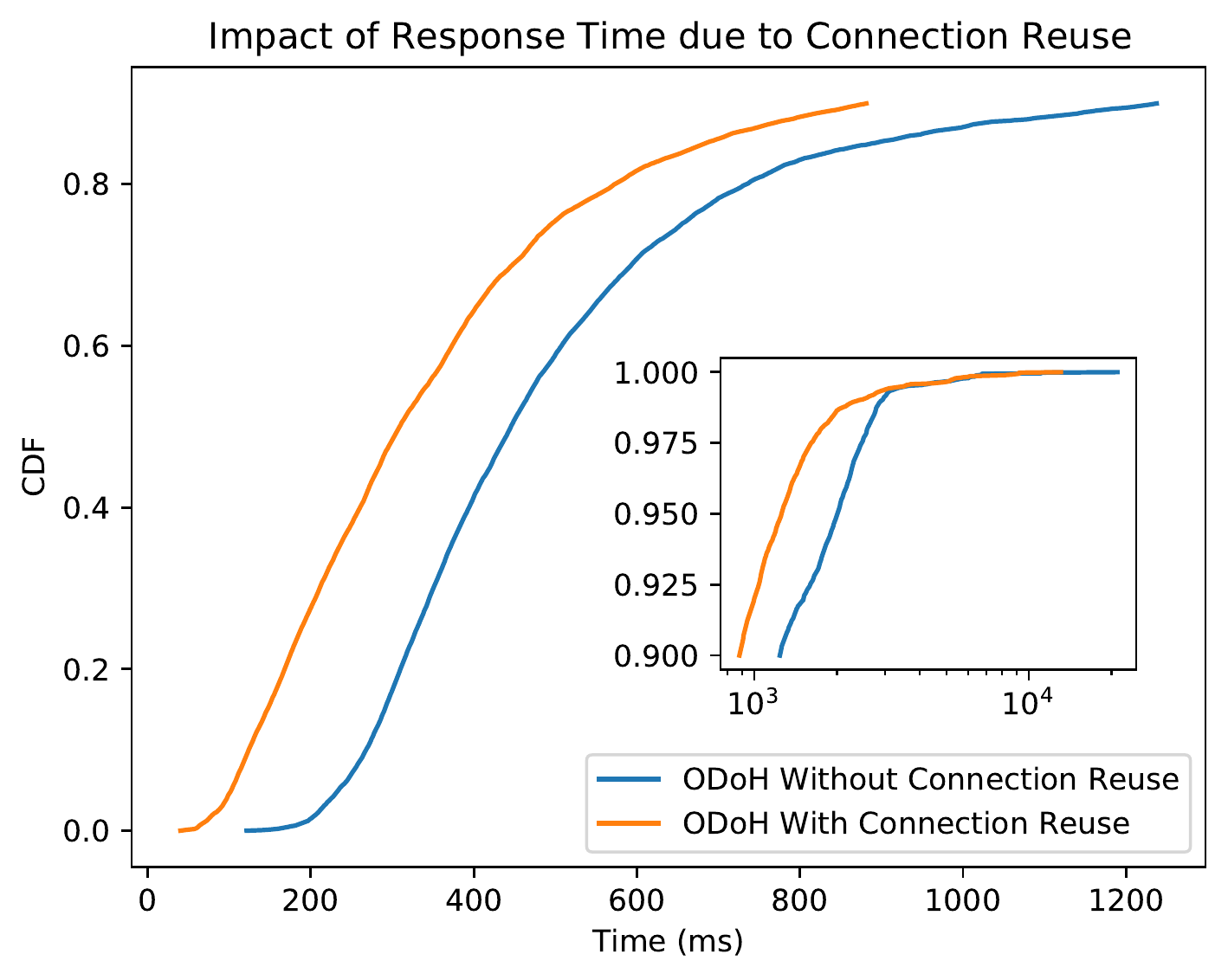}
	\caption{Impact of HTTPS connection reuse on ODoH}
	\label{fig:odoh_connection_reuse}
\end{figure}

Connection reuse is an optimization that enables clients to establish an HTTPS connection once per session, so that multiple queries can be executed in a single connection (rather than be forced to open a new connection for each query).
ODoH with and without connection reuse is shown in Figure~\ref{fig:odoh_connection_reuse}, using lowest latency proxy-target pair.
The average response time when clients use ODoH with persistent connections is 305.11 ms compared to 445.50 ms when the connection is not reused, a difference of 46.0\%.
We further discuss any limitations due to https connection reuse in Section~\ref{sec:limitations}.

\subsection{Impact of Service Co-location}
\label{ssec:colocation_impact}

In the measurements performed to this point, oblivious targets and resolvers are hosted individually and are physically separated.
This was to evaluate the penalty that is incurred by injecting the additional network hop. In a production environment, we envision popular DoH providers to enable support for \texttt{application/oblivious-dns-message} content types for DNS queries in their resolvers. This enables the target and resolver to be co-located, reducing network costs.
This penalty could be further reduced by enabling the target to execute 
a cache lookup for the query or communication with nameservers directly.

\begin{figure}[t!]
	\centering
	\includegraphics[width=0.8\columnwidth]{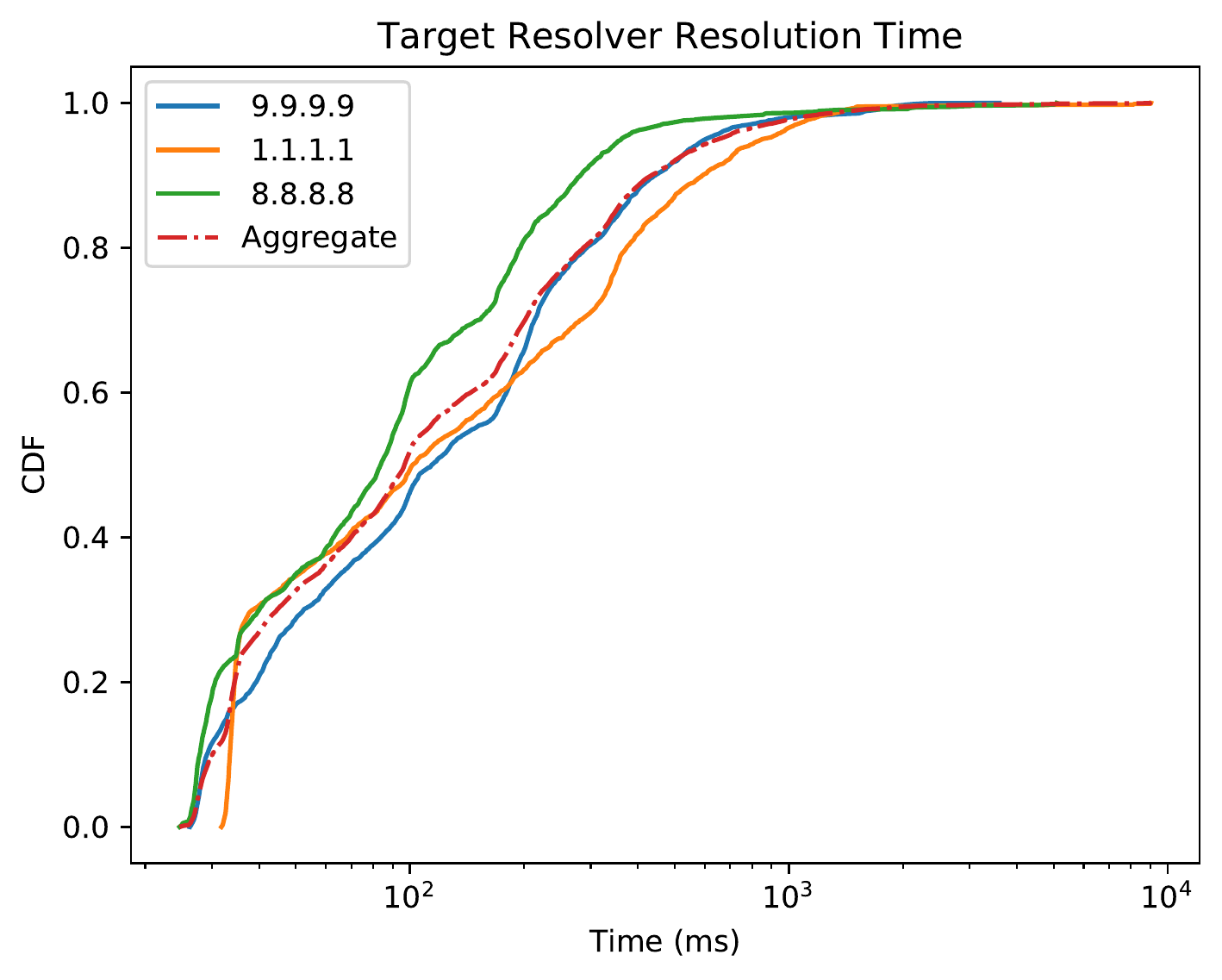}
	\caption{Time taken to resolve DoH query from Target instance to a chosen resolver (Target located in Google Cloud)}
	\label{fig:server_target_resolver_time}
\end{figure}

Using a daily production load from a popular open recursive DNS resolver~\cite{anon} we find that 68.6873\% of the queries result in a cache hit, precluding the need for further connections or communications. The result is a 99th percentile of 1ms for cache lookups.
Figure~\ref{fig:server_target_resolver_time} shows the time taken for oblivious targets to obtain an response from different recursive resolvers. This estimates the lower bound of the response time if the oblivious target was co-located with the resolver.
Co-location reduces the time between query encryption and answer decryption to 1ms for 68.6873\% of the queries due to the cache hit needed to retrieve the query's response. 
A cache miss incurs an
average response time of 120ms (at the 95th percentile) to communicate with other name servers, based on the same workload.
This results in an estimated lower bound of response time due to the usage of ODoH to be \textit{228.52} ms on an average as indicated by the dashed line in Figure~\ref{fig:all_protocols_compare}. Co-location of the oblivious target with the resolver will result in the best case results because of reduced network requests. The co-location and usage of ODoH will increase the DNS response times over DoH by an average of 56.3\%, instead of the experimental results of 103.2\%. This is reinforced by Figure~\ref{fig:server_target_resolver_time}, where Google's 8.8.8.8 response times are fastest to our \odoh targets that sit within Google's infrastructure.

\subsection{Impact of Network Type}
\label{ssec:odoh_network_impact}

\begin{figure}[t!]
	\centering
	\includegraphics[width=0.8\columnwidth]{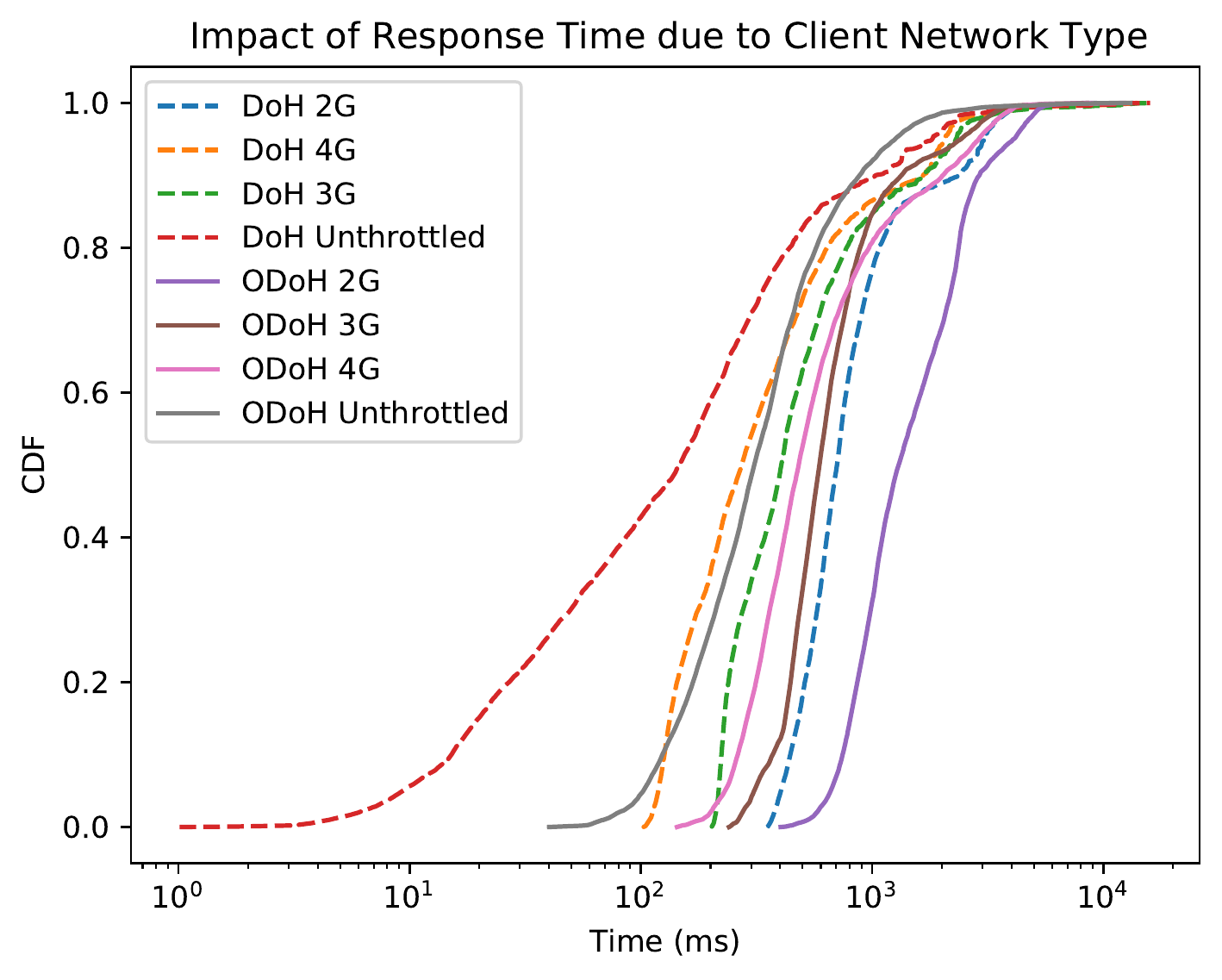}
	\caption{Time to resolve a \odoh query over different network conditions. The figure shows the comparison of the usage of ODOH and DOH  due to various network types.}
	\label{fig:odoh_throttle_impact}
\end{figure}

Borgolte \textit{et al.} indicated in their critique that the usage of DoH significantly impacted users with poorer network connectivity~\cite{borgolte2019dns}. We emulate various network conditions to understand their effect on ODoH.
In addition to the experimental setup
described in Section~\ref{sec:measurement_methodology}, we use the \texttt{qdisc} scheduler 
to configure and throttle egress bandwidth to 0.56 Mbps with 350 ms latency for 2G, 1.25 Mbps with 200 ms latency for 3G, and 12 Mbps with 100 ms latency for 4G~\cite{stanic2001tc, Downloadspeeds}. 
Our findings are presented in Figure~\ref{fig:odoh_throttle_impact}. We find that compared to unthrottled ODoH's average response time of 305.11 ms, the response times increase by 57.6\%, 94.0\% and 326\% for 4G, 3G and 2G network connections. We compare 
\odoh against DoH under the same conditions and find the response times are 76\%, 37\%, and 84\% higher compared to DoH in the same 4G, 3G and 2G networks, respectively. 


\subsection{Page Load Impact Due to ODOH}
\label{ssec:page_load_impact}

\begin{figure}[t!]
	\centering
	\includegraphics[width=0.9\columnwidth]{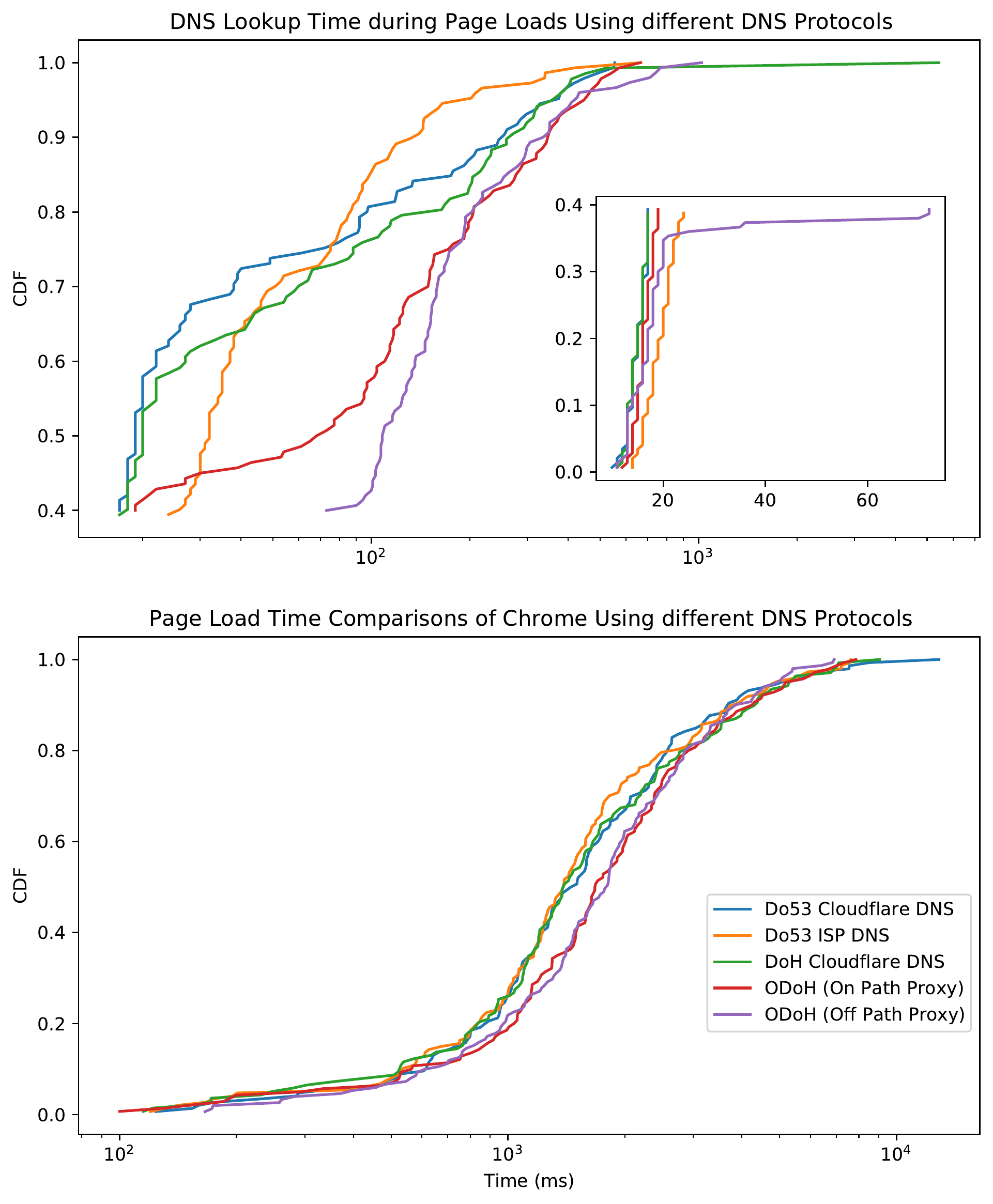}
	\caption{Impact of page load time due to usage of ODOH compared to other DNS protocols}
	\label{fig:plt_impact_odoh}
\end{figure}

The measurement of DoH to DNS performed by B\"{o}ttger \textit{et al.} concluded that
the performance penalty of DNS-over-HTTPS over clear text DNS is absorbed into, and has no significant perceived effect, on 
page load times~\cite{bottger2019empirical}.
We similarly evaluate page load times with ODoH. We replace the system DoH stub resolver in a popular open source DNS proxy and visit a set of 500 randomly chosen websites chosen from the top 2000 of the Tranco dataset~\cite{pochat2018tranco}. The measurements were completed on a university wireless network with a 200 Mbps download and 8 Mbps upload.
We measured page load times across five configurations: 
(1) the university's default DHCP-supplied DNS; 
(2) Do53 to Cloudflare's 1.1.1.1;
(3) DoH to 1.1.1.1;
(4) \odoh with proxy on the traceroute path to a target hosted on Google Cloud;
(5) \odoh with a proxy off-path to the same target hosted on Google Cloud.
The client stub resolver reuses connections where possible to avoid additional TCP and TLS handshakes. Page load times are measured using Selenium with the Chrome driver and the \texttt{window.performance} API. A new profile is created for each experimental run to remove caching effects.
Additionally, the local DNS stub resolver cache is flushed for each configuration. We calculate the page load time from information returned by \texttt{loadEventEnd}, and measure the DNS lookup times as the difference between \texttt{domainLookupEnd} and \texttt{domainLookupStart} events recorded by the browser. This is recommended by the W3C standard for page navigation time and perceived page load time~\cite{w3navigation}.

We present our results on the impact of page load time due to the change in the DNS protocol in Figure~\ref{fig:plt_impact_odoh}. The average page load time when using cloudflare DNS in clear-text UDP packets is 1325.0 ms. We consider the clear-text UDP result with Cloudflare as the baseline and find that the average page load times reduce by 0.4\% and 0.8\% when using the ISP issued DNS resolver and DoH to Cloudflare resulting in 1319.0 ms and 1314.0 ms respectively. The usage of ODoH with the on path proxy increases the page load times on an average by 21.2\% compared to the clear-text baseline usage to 1607.0 ms. The usage of ODoH with a proxy-target pair chosen where the proxy is not on the network path to the target increases the page load time by 25.5\% to 1663.0 ms. At the 95th percentile, we notice that both the DoH and ODoH protocols perform better than the baseline clear-text and UDP based DNS protocol by 23.6\% and 14\% respectively resulting in page load times of 5272.24 ms and 5692.29 ms when compared to 6516.75 ms with Cloudflare DNS and 5772.49 ms with the ISP provided DNS resolver. The difference between each ODoH variant and DoH (and Do53) can be attributed to the network path topology differences. For example, at the 70th percentile, the difference between ODoH (with an on-path proxy) and DoH is approximately 88.5ms. This difference consists of the DoH resolution time and the client-proxy and proxy-target RTTs (approximately 11ms and 1ms, respectively). 

Our results indicate that the client choice of proxy and targets strongly impacts the performance of the ODoH protocol and the corresponding page load time measurements. Additionally, we show that co-location of the oblivious target services with the recursive resolver will result in improved performance. We show in our page load time measurements that despite the lack of co-location of the target and the resolver services, there is a minimal impact on the perceived page load time for the user and argue that co-location will improve the performance of the ODoH protocol. We further show that ODoH can perform comparably (with minimal tradeoff) to DoH while providing both security and privacy guarantees to the user making it a practical replacement for clear-text DNS protocol usage.


\section{Limitations}
\label{sec:limitations}

\textbf{Collusion:} The ODoH protocol provides privacy guarantees when the oblivious target/resolver and the oblivious proxy operate as independent bodies and do not collude. Collusion among the proxy and the target operators can result in deanonymization of the client request and results in a loss of privacy for the client using the protocol. Similarly, the current protocol assumes a threat model where an adversary monitoring the network and being able to circumvent the channel encryption can only do so on a single network channel i.e. between two parties (client - proxy or proxy - target or client - target) with different adversaries monitoring the traffic between different parties. The privacy guarantees no longer hold if the same adversary is able to overcome https guarantees and monitor the traffic between the client and the proxy, and between the proxy and the target. However, in practice, the usage of https disallows adversaries from cryptographically circumventing the channel encryption to read the data being transferred. Adversaries in this setting can creating a mapping between the encrypted query and the encrypted response obtained from the target. Practical attacks on TLS which previously used the RC4 algorithm (CVE-2013-2566) indicated that a determined adversary needs to capture billions of connections which all contain the same text~\cite{NVD-CVERC4, aifardan2013security}. TLS1.2+ introduces the support for various authenticated encryption ciphers such as the AES cipher suites which are widely used in tools like Chrome for which no known practical attacks currently exist. We therefore consider risks due to colluding monitoring adversaries to be minimal.

\textbf{HTTPS Reuse:} In our experimental evaluations, the client stub resolvers try and reuse the https connection for sending different requests in an effort to improve performance and prevent performing the TCP and TLS handshakes with the proxy. The reuse of the session key for multiple queries could have some impact on the privacy of the user. However despite such connection reuse, no sensitive information regarding the query or the response in the encrypted or clear-text form is leaked. Some clients could prefer to explicitly disable connection reuse and create new https connections for every query to the proxy similar to DNSCrypt users. Doing so would provide the same privacy guarantees but have a considerable impact on performance i.e. time to resolve a query and page load times due to additional time for the TCP and TLS handshakes.

\textbf{DDoS Prevention:} The ODoH protocol prevents resolvers from using IP addresses to throttle malicious clients attempting to perform denial of service (DoS) attacks and moves the responsibility for the prevention of such attacks to the oblivious proxy which is receiving the queries. Proxies need to implement rate limiting procedures to prevent clients from performing a very large number of DNS requests which could be considered as an abuse of the system. However, malicious clients can issue these requests to different proxies which are forced to send the query to the targets. A malicious client could also encrypt a malformed DNS queries or check for non-existant domains in an effort to make the resolver spend compute cycles in decryption of the message or spend time communicating with various name servers~\cite{brownlee2001dns}.

\textbf{Key Reuse:} The ODoH protocol theoretically allows clients to choose the symmetric key corresponding to the authenicated encryption scheme being used to retrieve the encrypted answer from the oblivious target. This allows clients to use the same secret key for multiple queries allowing the target resolvers to identify the set of queries made by the same client. While this leakage of information does not leak the client IP address, it becomes possible for resolvers to group the queries together as that from the same client. Correct implementation of the protocol in the client stub resolver uses a (secure) cryptographically strong random number generator (RNG) to create the secret key preventing any possible reuse.

\textbf{PLT Triggers:} For the page load time measurements we perform in section~\ref{ssec:page_load_impact}, we use the \texttt{loadEventEnd} event which is triggered after all the sub resources of the page have been obtained. This result presented is the worst case performance compared to other metrics such as First Contentful Paint (FCP), First Meaningful Paint (FMP), and Time to Interactive (TTI) metrics which could be used for page load times. We believe that the implementation of ODoH with colocation of the target and the resolver in addition to the existing usage of DNS resolvers like Cloudflare, Akamai or Google DNS will result in better page load and DNS response times.


\section{Discussion}
\label{sec:discussion}

The results presented in Section~\ref{sec:results} indicate that ODoH can improve privacy of DNS queries while maintaining the security guarantees offered by DoH without significant performance penalties. Similar to DoH, real user privacy impacts with ODoH can only be achieved with significant adoption of the protocol. In this discussion, we present an adoption roadmap and steps necessary to make wide adoption reality. Additionally, we document attacks against ODoH, address some ethical and regulatory concerns with the usage of ODoH, and present ODoH as a solution to resolve criticisms of DoH~\cite{borgolte2019dns}.

\subsection{Adoption Roadmap}
\label{ssec:roadmap}
We believe that standardization of ODoH will result in an  increased adoption of the protocol. There has been increasing adoption of DoH post standardization due to support from operating system providers in Windows and iOS~\cite{AppleWWDCDoH, MicrosoftDoHInsider}. Applications, like Firefox and Chrome, have recently integrated support for DoH to specific trusted recursive resolvers~\cite{mozillaTRRProgram}. Developers of these platforms would largely prefer to avoid legal contracts and instead technologically guarantee privacy while providing their customers with the freedom to choose the resolver with the fastest response times. 

\textbf{Target Integration into DNS Resolvers:}
\label{sssec:dns_resolvers}
As adoption of third party open DNS resolvers like Cloudflare, Google, NextDNS, and Quad9 continues to grow, it would make sense for global anycast network operators offering DNS services guaranteeing user privacy to add support for ODoH in their resolvers. This includes a test domain like \texttt{odoh.test} with the corresponding DNSSEC signed HPKE public key information to be registered in the resolver for client discovery. In addition, the resolver should include support on the DoH API interface for ODoH messages with content type \texttt{application/oblivious-dns-message}, respecting the DNS wireformat, and introduce the services to decrypt a query and encrypt the answer. Support for HPKE is available in popular cryptographic libraries like BoringSSL and in language specific libraries written in go or rust~\cite{hpkeboringssl}. 

\textbf{Client Stub Resolver Modifications:}
\label{sssec:client_modifications}
Clients can use ODoH in multiple ways. Support for the protocol could be implemented directly by the OS's system stub resolver. Alternatively, support for the protocol could be provided by a DoH local proxy dns service which users install on their devices like cloudflared, dnscrypt-proxy, Acrylic, the experimental DoH-Proxy from Facebook, or applications like BraveDNS on Android devices~\cite{bravedns, cloudflared, facebookdohproxy}. Additional support for ODoH should require minimal changes to these tools, which already support DoH. We implemented the protocol in one such popular open source tool and used it to perform page load time measurements. Browsers like Firefox, and Chrome could include support for ODoH from the browser itself similar to DoH, preventing the need for users to run an system resolver which supports the protocol.

\textbf{Proxy Operators:}
\label{sssec:proxy_operators}
We imagine every DNS resolver operator with a CDN network, like Google, Akamai, Cloudflare, large cloud providers, and ISPs, could run ODoH proxy services. Additionally, the integration of ODoH into tools like DNSCrypt proxy could result in an increased adoption of the protocol. We expect these operators to operate oblivious proxies on the network for altruistic reasons, with a common shared goal of protecting user privacy similar to users running Tor relays. Since the usage of a proxy on the network path improves performance, we envision ISPs would want to provide proxy services. However, these proxies could be contrary to some ISP business models as they prohibit targeted DNS advertising. We believe that this may be a tractable concern given recent efforts by Mozilla in onboarding Comcast, a large ISP in the US, to their Trusted Recursive Resolver list~\cite{mozillacomcast20}. This may be due to DNS resolver operators shunning commercialization of DNS results due to concerns of regulatory backlash~\cite{borgolte2019dns, weaver2011redirecting}.

\subsection{Potential Attacks}
\label{ssec:potential_attacks}

\textbf{Association Attacks:} ODoH requires that the client stub resolver choose the proxy-target pair which belong to different non-colluding organizations to avoid the possibility of being able to associate an ODoH query to the client identity/IP address on the network. However, despite this configuration, there are a few scenarios where it becomes possible to create such associations between the client IP address and the request. If a client stub resolver chooses a target being run by the ISP of the client, it becomes possible for the ISP to identify the users' ODoH query proxied to it based on the immediate connections to the IP address returned in the response going through the ISP network. Similar risks of association do exist when the oblivious proxy is operated by the ISP of the client. The ISP can monitor the connections being initiated by the client immediately after the response from the target is obtained and sent to the client stub resolver. The ISP can make reasonable deductions given the client IP address and the destination IP requested to probabilistically guess the content of the encrypted query and tie the query to the client. However, the usage of ODoH in these scenarios increases the effort need to be put in by ISPs to track clients. Probabilistically correlating a query to the client IP address based on fingerprinting connection traffic does not increase the privacy risk than that posed by standard DoH queries.


\textbf{False Component Attacks:} We also consider the possibility of non-colluding attackers behaving as the components of ODoH. An adversary who behaves as a rogue oblivious target could log all the requests made and use this information to create encrypted messages from plaintext messages to create a chosen plaintext attack (CPA) and a ciphertext attack (CCA). 
An adversary behaving as a target logging the requests and responses could be legally compelled to provide the logs. ODOH prevents the ability to associate client IP addresses to the queries which were decrypted and resolved. An adversary behaving as an oblivious proxy could (1) drop the request altogether and prevent the client stub resolver from accessing DNS services, or (2) amplify the network traffic by creating spurious traffic to target resolvers which were not initiated by a client. This results in the adversary attempt at forcing the target to perform useless computation to attempt and decrypt a (potentially malformed) encrypted query. The proxy instance could also behave like a client and initiate encrypted DNS queries which are malformed. It is however possible in a production network to monitor and identify distributed denial of service (DDoS) or bot behavior through request amplification at the oblivious proxy and could be rate limited~\cite{CloudflareBots, DataDome, ClickGuard}. The proxy dropping client stub resolver packets can be identified by the client resolver using timeouts and could be configured to automatically use another proxy instance. Clients initiating spurious traffic as an attempt to overload the system isn't in scope of this work and poses the same risk as malformed cleartext UDP queries to the resolver.

\subsection{Ethical Implications}
\label{ssec:ethical_implications}
Encrypted DNS protocols pose significant challenges to enterprise system administrators who prefer to use clear-text DNS and parental controls which protect users inside the network from harmful content. Encrypted variants of DNS like DoH and ODoH make this difficult and required network operators to block outgoing network connections to a blacklist of constantly updated IP addresses. In these situations, we recommend enterprises which trust their network to use DoH queries to their enterprise DNS resolver but use ODoH for any outgoing connections from the organizations' gateway. This protects any potentially sensitive organization web traffic information from being recognized by other DNS name servers. Techniques to protect Internet traffic while maintaining local control over DNS can also be implemented in a home network with tools like PiHole which act as an Internet gateway for all web traffic from the home~\cite{Pihole}. Parental controls could be implemented directly on the PiHole device, allowing parents the ability to control searching patterns without DNS blocking.

ODoH is not intended to be a censorship-bypassing mechanism, but it does pose a potential problem for some current Internet filtering and blocking mechanisms. In some countries, there are guidances issued to ISPs to limit access to content such as child pornography or other illegal content. Many ISPs use DNS injection techniques to prevent their users from accessing such information~\cite{spence2005pennsylvania}. However, with DoH or ODoH and the usage of third party DNS resolvers supporting the protocol, ISPs lose visibility into DNS. This may cause them to choose to use other methods to comply to regulations, such as blocking connections to IP addresses of illegal content.

All anonymization techniques, irrespective of the protocol, can result in committed users circumventing the filtering guidelines of their ISP or region. However, ODoH providers in a given country can continue to abide by their legal regulations by limiting the inbound traffic to that from proxies hosted in countries with similar filtering guidelines. This may natually limit the usage of ODoH proxies by parties with which the client and/or resolver have no prior established relationship.

\section{Conclusion \& Future Work}
\label{sec:conclusion_future_work}
Clear-text DNS queries continue to be the most popular way in which users communicate with DNS ecosystem. In this paper, we perform the first implementation, deployment, and evaluation of an IETF-proposed  practical privacy enhancing technique called ODoH, with the aim to provide both security and privacy guarantees for users' DNS traffic by encrypting both the message and transport~\cite{odoh-draft}. We perform global measurements of the DNS response times, page load times and identify the various causes impacting performance of the protocol. We show that the proposed protocol is feasible and a practical replacement for DoH, improving user privacy and achieving minimal performance losses. We conclude with the limitations of our study, possible ethical implications, and sketch an adoption roadmap.

\bibliographystyle{plain}
\bibliography{references}

\begin{thebibliography}{10}

\bibitem{ClickGuard}
Clickguard | google ads click fraud detection \& protection software.
\newblock \url{https://www.clickguard.com/}.
\newblock (Accessed on 09/15/2020).

\bibitem{DataDome}
Datadome: Bot detection, protection and mitigation solution.
\newblock \url{https://datadome.co/bot-protection/}.
\newblock (Accessed on 09/15/2020).

\bibitem{DNSqueryAveragePiHole}
Dns query average : pihole.
\newblock
  \url{https://www.reddit.com/r/pihole/comments/a8ngnu/dns_query_average/}, 12
  2018.
\newblock (Accessed on 09/15/2020).

\bibitem{aifardan2013security}
N~Aifardan, D~Bernstein, K~Paterson, B~Poettering, and J~Schuldt.
\newblock On the security of rc4 in tls and wpa.
\newblock In {\em USENIX Security}, 2013.

\bibitem{anon}
Anonymous.
\newblock Anonymized for review, 17 Sept 2020.

\bibitem{ChromiumDoHBlog}
Kenji Baheux.
\newblock Chromium blog: A safer and more private browsing experience with
  secure dns.
\newblock
  \url{https://blog.chromium.org/2020/05/a-safer-and-more-private-browsing-DoH.html},
  05 2020.
\newblock (Accessed on 09/15/2020).

\bibitem{barnes2019hybrid}
Richard Barnes, Karthikeyan Bhargavan, Benjamin Lipp, and Christopher~A. Wood.
\newblock {Hybrid Public Key Encryption}.
\newblock Internet-Draft draft-irtf-cfrg-hpke-05, Internet Engineering Task
  Force, July 2020.
\newblock Work in Progress.

\bibitem{bernstein2009dnscurve}
Daniel~J Bernstein.
\newblock Dnscurve: Usable security for dns.
\newblock {\em dnscurve. org}, 4, 2009.

\bibitem{berthold2001web}
Oliver Berthold, Hannes Federrath, and Stefan K{\"o}psell.
\newblock Web mixes: A system for anonymous and unobservable internet access.
\newblock In {\em Designing privacy enhancing technologies}, pages 115--129.
  Springer, 2001.

\bibitem{borgolte2019dns}
Kevin Borgolte, Tithi Chattopadhyay, Nick Feamster, Mihir Kshirsagar, Jordan
  Holland, Austin Hounsel, and Paul Schmitt.
\newblock How dns over https is reshaping privacy, performance, and policy in
  the internet ecosystem.
\newblock {\em Performance, and Policy in the Internet Ecosystem (July 27,
  2019)}, 2019.

\bibitem{bortzmeyer2015dns}
Stephane Bortzmeyer.
\newblock Dns privacy considerations.
\newblock {\em Work in Progress, draft-ietf-dprive-problem-statement-06}, 1,
  2015.

\bibitem{bortzmeyer2016dns}
Stephane Bortzmeyer.
\newblock Dns query name minimisation to improve privacy.
\newblock {\em RFC7816}, 2016.

\bibitem{bottger2019empirical}
Timm B{\"o}ttger, Felix Cuadrado, Gianni Antichi, Eder~Le{\~a}o Fernandes,
  Gareth Tyson, Ignacio Castro, and Steve Uhlig.
\newblock An empirical study of the cost of dns-over-https.
\newblock In {\em Proceedings of the Internet Measurement Conference}, pages
  15--21, 2019.

\bibitem{bravedns}
BraveDNS.
\newblock bravedns - a fast, secure, configurable, private dns + firewall for
  android.
\newblock \url{https://www.bravedns.com/}.
\newblock (Accessed on 09/16/2020).

\bibitem{brownlee2001dns}
Nevil Brownlee, Kimberly~C Claffy, and Evi Nemeth.
\newblock Dns measurements at a root server.
\newblock In {\em GLOBECOM'01. IEEE Global Telecommunications Conference (Cat.
  No. 01CH37270)}, volume~3, pages 1672--1676. IEEE, 2001.

\bibitem{castillo2009evaluation}
Sergio Castillo-Perez and Joaquin Garcia-Alfaro.
\newblock Evaluation of two privacy-preserving protocols for the dns.
\newblock In {\em 2009 Sixth International Conference on Information
  Technology: New Generations}, pages 411--416. IEEE, 2009.

\bibitem{cleanbrowsing}
Cleanbrowsing.
\newblock Cleanbrowsing. a dns-based, privacy-first, content filtering service.
  blocks malicious, pornographic, gambling and other simmilar site categories
  with a fast and reliable dns. protect your home, school, library and company.
\newblock \url{https://cleanbrowsing.org/}.
\newblock (Accessed on 09/15/2020).

\bibitem{gcecloud}
Google Cloud.
\newblock Google compute engine - machine types.
\newblock \url{https://cloud.google.com/compute/docs/machine-types}.
\newblock (Accessed on 09/16/2020).

\bibitem{CloudflareBots}
Cloudflare.
\newblock Cloudflare bot management.
\newblock \url{https://www.cloudflare.com/products/bot-management/}.
\newblock (Accessed on 09/15/2020).

\bibitem{CloudflareWorkers}
Cloudflare.
\newblock Cloudflare workers®.
\newblock \url{https://workers.cloudflare.com/}.
\newblock (Accessed on 09/15/2020).

\bibitem{DNSoverHTTPSCloudflare}
Cloudflare.
\newblock Dns over https | cloudflare developer docs.
\newblock \url{https://developers.cloudflare.com/1.1.1.1/dns-over-https}.
\newblock (Accessed on 09/15/2020).

\bibitem{DNSoverTorCloudflare}
Cloudflare.
\newblock Dns over tor | cloudflare developer docs.
\newblock
  \url{https://developers.cloudflare.com/1.1.1.1/fun-stuff/dns-over-tor/}.
\newblock (Accessed on 09/15/2020).

\bibitem{cloudflared}
Cloudflare.
\newblock Argo tunnel client.
\newblock \url{https://github.com/cloudflare/cloudflared}, 2020.

\bibitem{federal2016protecting}
Federal~Communications Commission et~al.
\newblock Protecting the privacy of customers of broadband and other
  telecommunications services.
\newblock {\em Report and order, Washington, DC, USA, Nov}, 2016.

\bibitem{dagon2008increased}
David Dagon, Manos Antonakakis, Paul Vixie, Tatuya Jinmei, and Wenke Lee.
\newblock Increased dns forgery resistance through 0x20-bit encoding: security
  via leet queries.
\newblock In {\em Proceedings of the 15th ACM conference on Computer and
  communications security}, pages 211--222, 2008.

\bibitem{FirefoxDoH}
Selena Deckelmann.
\newblock Firefox continues push to bring dns over https by default for us
  users - the mozilla blog.
\newblock
  \url{https://blog.mozilla.org/blog/2020/02/25/firefox-continues-push-to-bring-dns-over-https-by-default-for-us-users/},
  02 2020.
\newblock (Accessed on 09/15/2020).

\bibitem{AnonymizedDNSCrypt}
Frank Denis.
\newblock Anonymized {DNSCrypt} specification.
\newblock
  \url{https://github.com/DNSCrypt/dnscrypt-protocol/blob/master/ANONYMIZED-DNSCRYPT.txt},
  06 2020.
\newblock (Accessed on 09/15/2020).

\bibitem{DNSCryptProxyImpl}
Frank Denis and Contributors.
\newblock A flexible dns proxy, with support for encrypted dns protocols.
\newblock \url{https://github.com/DNSCrypt/dnscrypt-proxy/}.
\newblock (Accessed on 09/17/2020).

\bibitem{DNSCrypt}
Frank Denis and Yecheng Fu.
\newblock Dnscrypt, 2015.

\bibitem{AppleDNSProxy}
Apple Developer.
\newblock Dns proxy provider | apple developer documentation.
\newblock
  \url{https://developer.apple.com/documentation/networkextension/dns_proxy_provider}.
\newblock (Accessed on 09/15/2020).

\bibitem{AppleWWDCDoH}
Apple Developer.
\newblock Enable encrypted dns - wwdc 2020.
\newblock \url{https://developer.apple.com/videos/play/wwdc2020/10047/}.
\newblock (Accessed on 09/15/2020).

\bibitem{AndroidDNSDoT}
Android Developers.
\newblock Dns over tls support in android p developer preview.
\newblock
  \url{https://android-developers.googleblog.com/2018/04/dns-over-tls-support-in-android-p.html},
  04 2018.
\newblock (Accessed on 09/15/2020).

\bibitem{cloudflareprivacy}
Cloudflare~Developer Docs.
\newblock 1.1.1.1 public dns resolver - cloudflare's commitment to privacy.
\newblock
  \url{https://developers.cloudflare.com/1.1.1.1/privacy/public-dns-resolver/}.
\newblock (Accessed on 09/15/2020).

\bibitem{dolev1983security}
Danny Dolev and Andrew Yao.
\newblock On the security of public key protocols.
\newblock {\em IEEE Transactions on information theory}, 29(2):198--208, 1983.

\bibitem{facebookdohproxy}
Facebook.
\newblock Dns over https proxy | facebook.
\newblock \url{https://github.com/facebookexperimental/doh-proxy}, 2020.

\bibitem{federrath2011privacy}
Hannes Federrath, Karl-Peter Fuchs, Dominik Herrmann, and Christopher Piosecny.
\newblock {Privacy-preserving DNS: analysis of broadcast, range queries and
  mix-based protection methods}.
\newblock In {\em European Symposium on Research in Computer Security}, pages
  665--683. Springer, 2011.

\bibitem{DNSoverHTTPSGoogle}
Google.
\newblock Dns-over-https (doh) | public dns | google developers.
\newblock \url{https://developers.google.com/speed/public-dns/docs/doh}.
\newblock (Accessed on 09/15/2020).

\bibitem{hpkeboringssl}
Google.
\newblock crypto/hpke - boringssl - git at google.
\newblock
  \url{https://boringssl.googlesource.com/boringssl/+/refs/heads/master/crypto/hpke/},
  07 2020.
\newblock (Accessed on 09/17/2020).

\bibitem{CloudflareDNSAudit}
John Graham-Cumming.
\newblock Announcing the results of the 1.1.1.1 public dns resolver privacy
  examination.
\newblock
  \url{https://blog.cloudflare.com/announcing-the-results-of-the-1-1-1-1-public-dns-resolver-privacy-examination/},
  03 2020.
\newblock (Accessed on 09/15/2020).

\bibitem{greschbach2016effect}
Benjamin Greschbach, Tobias Pulls, Laura~M Roberts, Philipp Winter, and Nick
  Feamster.
\newblock The effect of dns on tor's anonymity.
\newblock {\em arXiv preprint arXiv:1609.08187}, 2016.

\bibitem{morecowbell}
Christian Grothoff, Matthias Wachs, Monika Ermert, and Jacob Appelbaum.
\newblock {NSA's} morecowbell: Knell for dns, 2015.

\bibitem{TorSeattleCapture}
Ansel Herz.
\newblock Judge who authorized police search of seattle privacy activists
  wasn't told they operate tor network.
\newblock
  \url{https://web.archive.org/web/20191210114929/https://www.thestranger.com/slog/2016/04/08/23914735/judge-who-authorized-police-search-of-seattle-privacy-activists-wasnt-told-they-operate-tor-network/},
  04 2016.
\newblock (Accessed on 09/15/2020).

\bibitem{hoffman2018dns}
Paul Hoffman and Patrick McManus.
\newblock Dns queries over https (doh).
\newblock {\em Internet Requests for Comments, IETF, RFC}, 8484, 2018.

\bibitem{Pihole}
Pi~Hole.
\newblock Pi-hole – a black hole for internet advertisements.
\newblock \url{https://pi-hole.net/}.
\newblock (Accessed on 09/16/2020).

\bibitem{hounsel2020measuring}
Austin Hounsel, Paul Schmitt, Kevin Borgolte, and Nick Feamster.
\newblock Measuring the performance of encrypted dns protocols from broadband
  access networks, 2020.

\bibitem{housley2011reducing}
Russ Housley, D~Crocker, and E~Burger.
\newblock Reducing the standards track to two maturity levels.
\newblock {\em IETF RFC 6410}, 2011.

\bibitem{hu2016specification}
Zi~Hu, Liang Zhu, John Heidemann, Allison Mankin, Duane Wessels, and Paul
  Hoffman.
\newblock Specification for dns over transport layer security (tls).
\newblock {\em IETF RFC7858, May}, 2016.

\bibitem{ISPAMozillaVillain19}
{ISPA UK}.
\newblock Ispa announces finalists for 2019 internet heroes and villains: Trump
  and mozilla lead the way as villain nominees - the internet service providers
  association.
\newblock
  \url{https://www.ispa.org.uk/ispa-announces-finalists-for-2019-internet-heroes-and-villains-trump-and-mozilla-lead-the-way-as-villain-nominees/},
  07 2019.
\newblock (Accessed on 09/15/2020).

\bibitem{AESGCMPerformance}
Franziskus Kiefer.
\newblock Improving aes-gcm performance - mozilla security blog.
\newblock
  \url{https://blog.mozilla.org/security/2017/09/29/improving-aes-gcm-performance/},
  09 2017.
\newblock (Accessed on 09/16/2020).

\bibitem{odoh-draft}
E.~Kinnear, P.~McManus, T.~Pauly, and C.~Wood.
\newblock Oblivious dns over https--ietf draft.
\newblock
  \url{https://tools.ietf.org/html/draft-pauly-dprive-oblivious-doh-01}, 2019.

\bibitem{MicrosoftDoHInsider}
Brandon {LeBlanc}.
\newblock Announcing windows 10 insider preview build 20185.
\newblock
  \url{https://blogs.windows.com/windows-insider/2020/08/05/announcing-windows-10-insider-preview-build-20185/},
  08 2020.
\newblock (Accessed on 09/15/2020).

\bibitem{Downloadspeeds}
Ken Lo.
\newblock Download speeds: Comparing 2g, 3g, 4g \& 5g mobile networks.
\newblock
  \url{https://kenstechtips.com/index.php/download-speeds-2g-3g-and-4g-actual-meaning},
  11 2018.
\newblock (Accessed on 09/16/2020).

\bibitem{lu2019end}
Chaoyi Lu, Baojun Liu, Zhou Li, Shuang Hao, Haixin Duan, Mingming Zhang,
  Chunying Leng, Ying Liu, Zaifeng Zhang, and Jianping Wu.
\newblock An end-to-end, large-scale measurement of dns-over-encryption: How
  far have we come?
\newblock In {\em Proceedings of the Internet Measurement Conference}, pages
  22--35, 2019.

\bibitem{TorEFFMisunderstanding}
Electronic Frontier~Foundation Marcia~Hoffmann.
\newblock Why ip addresses alone don't identify criminals.
\newblock
  \url{https://www.eff.org/deeplinks/2011/08/why-ip-addresses-alone-dont-identify-criminals},
  08 2011.
\newblock (Accessed on 09/15/2020).

\bibitem{mockapetris1987domain}
Paul Mockapetris et~al.
\newblock Domain names-implementation and specification.
\newblock 1987.

\bibitem{mozillacomcast20}
Mozilla.
\newblock Comcast’s xfinity internet service joins firefox’s trusted
  recursive resolver program - the mozilla blog.
\newblock
  \url{https://blog.mozilla.org/blog/2020/06/25/comcasts-xfinity-internet-service-joins-firefoxs-trusted-recursive-resolver-program/},
  06 2020.
\newblock (Accessed on 09/15/2020).

\bibitem{mozillaTRRProgram}
Mozilla.
\newblock Mozilla policy requirements for dns over https partners.
\newblock \url{https://wiki.mozilla.org/Security/DOH-resolver-policy}, 09 2020.
\newblock (Accessed on 09/15/2020).

\bibitem{DoHoTPractical}
Alec Muffett.
\newblock Dohot: making practical use of dns over https over tor.
\newblock \url{https://github.com/alecmuffett/dohot}, 07 2020.
\newblock (Accessed on 09/15/2020).

\bibitem{NVD-CVERC4}
{NIST}.
\newblock Nvd - cve-2013-2566.
\newblock \url{https://nvd.nist.gov/vuln/detail/CVE-2013-2566}, 03 2013.
\newblock (Accessed on 09/16/2020).

\bibitem{pochat2018tranco}
Victor~Le Pochat, Tom Van~Goethem, Samaneh Tajalizadehkhoob, Maciej
  Korczy{\'n}ski, and Wouter Joosen.
\newblock Tranco: A research-oriented top sites ranking hardened against
  manipulation.
\newblock {\em arXiv preprint arXiv:1806.01156}, 2018.

\bibitem{ChromeDoH}
Chromium Projects.
\newblock Dns over https (aka doh).
\newblock \url{https://www.chromium.org/developers/dns-over-https}.
\newblock (Accessed on 09/15/2020).

\bibitem{AnonDNSCryptImplementation}
{DNSCrypt} Proxy.
\newblock Anonymized dns wiki.
\newblock \url{https://github.com/DNSCrypt/dnscrypt-proxy/wiki/Anonymized-DNS}.
\newblock (Accessed on 09/15/2020).

\bibitem{pub1994security}
FIPS PUB.
\newblock Security requirements for cryptographic modules.
\newblock {\em FIPS PUB}, 140, 1994.

\bibitem{DNSoverHTTPSQuad9}
Quad9.
\newblock Doh support via quad9 dns - quad 9.
\newblock \url{https://www.quad9.net/doh-quad9-dns-servers/}.
\newblock (Accessed on 09/15/2020).

\bibitem{schmitt2019oblivious_full}
Paul Schmitt, Anne Edmundson, Allison Mankin, and Nick Feamster.
\newblock Oblivious dns: Practical privacy for dns queries.
\newblock {\em Proceedings on Privacy Enhancing Technologies},
  2019(2):228--244, 2019.

\bibitem{spence2005pennsylvania}
John~B Spence.
\newblock Pennsylvania and pornography.
\newblock {\em Available at SSRN 669861}, 2005.

\bibitem{stanic2001tc}
Milan~P Stanic.
\newblock Tc--traffic control.
\newblock {\em Linux QOS Control Tool}, 2001.

\bibitem{sundaresan2013web}
Srikanth Sundaresan, Nazanin Magharei, Nick Feamster, Renata Teixeira, and Sam
  Crawford.
\newblock Web performance bottlenecks in broadband access networks.
\newblock In {\em Proceedings of the ACM SIGMETRICS/international conference on
  Measurement and modeling of computer systems}, pages 383--384, 2013.

\bibitem{TorBanUAE}
TracBot.
\newblock Tor blocked in uae (\#25137) · issues · legacy / trac · gitlab.
\newblock \url{https://gitlab.torproject.org/legacy/trac/-/issues/25137}, 02
  2018.
\newblock (Accessed on 09/15/2020).

\bibitem{ISPView}
Upturn.
\newblock What isps can see.
\newblock \url{https://www.upturn.org/reports/2016/what-isps-can-see/}, 03
  2016.
\newblock (Accessed on 09/15/2020).

\bibitem{w3navigation}
Zhiheng Wang.
\newblock Navigation timing - world wide web consortium (w3c).
\newblock \url{https://www.w3.org/TR/navigation-timing/}, 12 2012.
\newblock (Accessed on 09/17/2020).

\bibitem{weaver2011redirecting}
Nicholas Weaver, Christian Kreibich, and Vern Paxson.
\newblock Redirecting dns for ads and profit.
\newblock {\em FOCI}, 2:2--3, 2011.

\bibitem{TorBanIran}
{Xynou, Maria, and Filasto, Artur\`{o}}.
\newblock Iran protests: Ooni data confirms censorship events (part 1) | ooni.
\newblock \url{https://ooni.org/post/2018-iran-protests/}.
\newblock (Accessed on 09/15/2020).

\bibitem{zhao2007analysis}
Fangming Zhao, Yoshiaki Hori, and Kouichi Sakurai.
\newblock Analysis of privacy disclosure in dns query.
\newblock In {\em 2007 International Conference on Multimedia and Ubiquitous
  Engineering (MUE'07)}, pages 952--957. IEEE, 2007.

\bibitem{zhao2007two}
Fangming Zhao, Yoshiaki Hori, and Kouichi Sakurai.
\newblock Two-servers pir based dns query scheme with privacy-preserving.
\newblock In {\em The 2007 International Conference on Intelligent Pervasive
  Computing (IPC 2007)}, pages 299--302. IEEE, 2007.

\bibitem{zhu2014t}
Liang Zhu, Zi~Hu, John Heidemann, Duane Wessels, Allison Mankin, and Nikita
  Somaiya.
\newblock T-dns: Connection-oriented dns to improve privacy and security.
\newblock {\em ACM SIGCOMM Computer Communication Review}, 44(4):379--380,
  2014.

\end{thebibliography}

\end{document}